\documentclass[prb,floatfix,twocolumn,showpacs,superscriptaddress]{revtex4}
\usepackage{graphicx,array,amssymb}
\usepackage{color}
\def\bea{\begin{eqnarray}}
\def\eea{\end{eqnarray}}
\def\be{\begin{equation}}
\def\ee{\end{equation}}

\begin{document}

\author{Matthieu Mambrini}

\affiliation{
  Laboratoire de Physique Th\'eorique, Universit\'e de Toulouse, 
  F-31062 Toulouse, France }

\author{Andreas L\"auchli}
\affiliation{
  Institut Romand de Recherche Num\'erique en Physique des Mat\'eriaux (IRRMA),
  CH-1015 Lausanne, Switzerland}
  
\author{Didier Poilblanc}

\affiliation{
  Laboratoire de Physique Th\'eorique, Universit\'e de Toulouse,
  F-31062 Toulouse, France }
\affiliation{
Institute of Theoretical Physics,
Ecole Polytechnique F\'ed\'erale de Lausanne,
BSP 720,
CH-1015 Lausanne,
Switzerland}

\author{Fr\'ed\'eric Mila}
\affiliation{
Institute of Theoretical Physics,
Ecole Polytechnique F\'ed\'erale de Lausanne,
BSP 720,
CH-1015 Lausanne,
Switzerland}

\date{\today}
\title{Plaquette valence bond crystal in the frustrated   
Heisenberg quantum antiferromagnet on the square lattice}

\pacs{75.10.-b, 75.10.Jm, 75.40.Mg}
\begin{abstract}

Using both exact diagonalizations and diagonalizations in a subset of short-range valence bond singlets, we address
the nature of the groundstate of the Heisenberg spin-$1/2$ antiferromagnet on the square lattice with
competing next-nearest and next-next-nearest neighbor antiferromagnetic couplings ($J_1{-}J_2{-}J_3$ model). A detailed comparison of
the two approaches reveals a region along the line $(J_2+J_3)/J_1=1/2$, where the description in terms of nearest-neighbor
singlet coverings is excellent, therefore providing evidence for a magnetically disordered region.
Furthermore a careful analysis of dimer-dimer correlation functions, dimer structure factors and plaquette-plaquette
correlation functions provides striking evidence for the presence of a plaquette valence bond crystal order in part of
the magnetically disordered region.

\end{abstract}
\maketitle

\section{Introduction}

Frustration can drive the low-energy physics of bidimensional Heisenberg quantum antiferromagnets
very far from conventional semi-classical N\'eel-like phases. In such a case, the breakdown of long range magnetic
order in the ground state leads the system to reorganize in a typical quantum state
where only local antiferromagnetic correlations are present, namely a superposition of short range valence
bond (SRVB) states. In this regime, the system opens a gap to the magnetic excitations and the SU(2) symmetry of
the Hamiltonian is restored. However, inside this general frame, the nature of the SRVB ground state (GS) can be very different
from one system to another. In the simplest scenario the spatial symmetry of the Hamiltonian can still be broken, leading to a valence bond
crystal phase (VBC) characterized by long range order in the dimer-dimer correlation function\cite{read_sachdev}. Alternatively, all
symmetries can be restored in a flat superposition of SRVB states to form a so-called spin liquid (SL).

Far from being purely academic, the precise determination of the GS nature is a crucial question in the context of quantum phase transitions\cite{sachdev_book}. For example,
the ``deconfined critical point'' (DCP) scenario has been proposed as a new class of criticality to describe the N\'eel to VBC transition\cite{senthil04,senthil05}. More importantly, the nature
of the magnetic background dramatically affects the holon/spinon (de)confinement properties of the corresponding doped systems. It is therefore believed to be a key ingredient to understand exotic metallic states.

In practice, it is often hard to fully characterize the type, from crystal to liquid, of a SRVB phase. In this respect, one of the most archetypal example of such a situation is the $J_1 - J_2$  Heisenberg $S=1/2$ antiferromagnet on the square lattice, where frustration is controlled by the next nearest neighbor interaction $J_2$. Despite many years of numerical and analytical efforts, no definitive picture emerged around the maximally frustrated 
point $J_2 /J_1 \sim 0.5$, where the magnetic order disappears. The main point of this article is to introduce a general framework to study this kind of highly frustrated antiferromagnet (the SRVB method) and to revisit the question on this specific model within an extended version of the Hamiltonian including a third neighbor $J_3$ interaction~:

\begin{equation}
    {\cal H} = J_1 \sum_{\langle i,j \rangle} {\bf S}_i . {\bf S}_j \\
    +   J_2 \sum_{\langle \langle i,j \rangle \rangle} {\bf S}_i . {\bf S}_j \\
    + J_3 \sum_{\langle \langle \langle i,j \rangle \rangle \rangle} {\bf S}_i . {\bf S}_j
\end{equation}

\begin{figure}
\centerline{\includegraphics*[angle=0,width=0.9\linewidth]{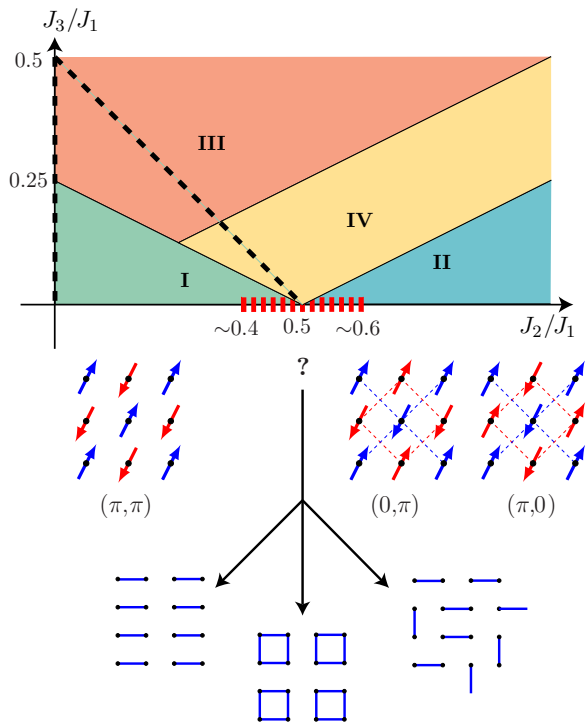}}
  \caption{\label{fig:phases}
(Color online) Classical phase diagram of the $J_1-J_2-J_3$ model : {\bf I} N\'eel $(\pi,\pi)$, {\bf II} Collinear $(0,\pi)$ and $(\pi,0)$,
{\bf III} Helicoidal $(q,q)$, {\bf IV} Helicoidal $(q,\pi)$. The snapshots refer to the quantum version of the $J_1-J_2$ model and in particular
the various possible scenarios in the magnetically disordered gapped (red dotted line) phase around $J_2/J_1 \sim 0.5$ : columnar VBC, plaquette VBC or spin liquid. VBC correlations are investigated in details along the black dashed lines in the present paper.}
\end{figure}

At a classical level~\cite{J1J2J3,Chubukov,Ferrer}, the effect of frustration and competition between $J_2/J_1$ and $J_3/J_1$
leads to four ordered phases described in figure \ref{fig:phases}. The effect of quantum fluctuations on this classical phase diagram is still
an open question. 
In the past 15 years, the situation has been somewhat
clarified for the pure $J_1-J_2$ model, 
 especially in a range of parameters
far from the maximally frustrated point $J_2 /J_1 \sim 0.5$.
For $J_3 =0 $ and $J_2/J_1 \lesssim 0.4$,
the classical $(\pi,\pi)$ N\'eel behavior
is essentially conserved~\cite{ChandraDoucot,j1j2} up to a small reduction of the staggered magnetization.
On the other hand, for $J_2/J_1 \gtrsim 0.6$
an {\it order by disorder} mechanism~\cite{ChandraColeman} selects two
collinear states at ${\bf q}=(\pi,0)$ and $(0,\pi)$. In the parameter
range where frustration is the largest,
$0.4 \lesssim J_2/J_1 \lesssim 0.6$,
the situation is much more involved.
Beside the fact that
many approaches (including spin-wave theory~\cite{ChandraDoucot},
exact diagonalizations~\cite{j1j2},
series-expansion~\cite{GelfandSinghHuse} and large-$N$
expansions~\cite{spinons}) have now firmly established that
quantum fluctuations destabilize the classical ordered ground
state and lead to a quantum disordered singlet ground state with a gap to the first magnetic excitation,
its precise nature is still controversial~:
a columnar valence bond crystal with both translational and rotational
broken symmetries~\cite{read_sachdev}, a plaquette state with no broken rotational
symmetry~\cite{plaquette} or even a spin-liquid with no broken
symmetry~\cite{CapriottiBecca} have been proposed (see figure \ref{fig:phases}).

For the $J_3 \neq 0$ case, as remarked by Ferrer~\cite{Ferrer},
the end point of the classical critical line $(J_2+2J_3)/J_1=1/2$
on the $J_3$ axis is substantially shifted to larger values of
$J_3$ when quantum fluctuations are switched on. For the pure
$J_1-J_3$ model, in this region of large frustration, a
non-classical (but still controversial) phase appears between the
N\'eel $(\pi,\pi)$ and the spiral $(q,q)$ phases~: a VBC columnar
state~\cite{Leung}, a spin-liquid~\cite{Capriotti} or a succession
of a VBC and $Z_2$ spin-liquid phases~\cite{CapriottiSachdev} have
been proposed. The complete phase diagram of the  $J_1-J_2-J_3$ quantum antiferromagnet is expected to be even richer.
Indeed, preliminary calculations~\cite{Figueirido} pointed towards an extended region with a quantum disordered state.

In this paper we investigate the maximally frustrated region of this phase
diagram $(J_2+J_3)/J_1 \sim 1/2$ (dashed line in figure \ref{fig:phases}) using both exact diagonalizations and 
a SRVB method which consists in diagonalizing the Hamiltonian in a subset
of singlets states that can be written in term of SRVB states.
In the first section, we introduce in details the method as a natural tool 
to study magnetically
disordered phases and discuss its advantages and limitations. In the 
second part,
we show numerical evidences for an extended non-magnetic phase around 
$(J_2+J_3)/J_1 \sim 1/2$.
In a third part we present calculations and finite size analysis of dimer-dimer
correlation functions and dimer structure factors that establishes the existence of 
a {\em s-wave plaquette ordered phase 
breaking only translational symmetry when $J_3 \geq J_2$ and $(J_2+J_3)/J_1 \sim 1/2$.}
This point is directly confirmed in the last part by an inspection of plaquette-plaquette correlations.
We conclude by emphasizing the interest of the $J_1-J_3$ model as an example of N\'eel to VBC quantum phase transition and discuss the 
implications of our results for the much debated $J_1-J_2$ model.

\section{SRVB method}

From a numerical point of view, investigating the low energy physics of 2d frustrated quantum 
antiferromagnets is a difficult problem. Among the three well known high precision and controlled methods,
two of them cannot be applied, at least for the moment~: Density Matrix Renormalization Group (DMRG) is only efficient in
one dimension and Quantum Monte Carlo (QMC) suffers from a severe sign problem on these systems. The third 
method, namely exact diagonalizations (ED), consists in a complete enumeration of the Hilbert space followed by an
iterative solving of the eigenproblem. The main advantages of this approach are (i) it is numerically exact, (ii) any observable
is accessible (iii) spatial symmetries can be fully taken into account thus providing momentum resolved results.
Unfortunately, the first step of the method faces the exponential growth
of the Hilbert space
with system size for finite available computing resources. 
Nevertheless, this method is still
widely and successfully used and is the source of 
many firmly established results.

However, if one compares highly frustrated quantum antiferromagnets 
to more conventional
unfrustrated ones (typically N\'eel like), a phenomenological review of known results
shows that
\begin{enumerate}
\item the role of the singlet sector is overwhelming at low energy due to the opening
of a singlet-triplet gap,
\item the breakdown of antiferromagnetic
long range order favors the emergence of local singlet patterns.
\end{enumerate}
In this respect, it is tempting to build a more specific approach taking into account these two points in order
to systematically reduce the Hilbert space to a relevant subset adapted to describe magnetically disordered singlet states.

\begin{figure}
\centerline{\includegraphics*[angle=0,width=0.9\linewidth]{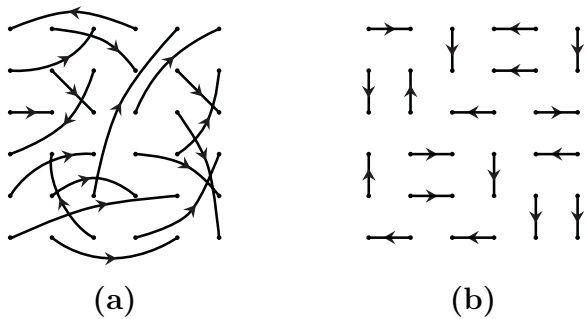}}
  \caption{\label{fig:srrvb}
(a) Arbitrary range VB state. (b) Nearest neighbor VB state (NNVB). The oriented bond between two sites $i$ and $j$
stands for $[i,j] = (1/\sqrt{2}) (\vert\uparrow_i \downarrow_j \rangle - \vert \downarrow_i \uparrow_j \rangle)$.}
\end{figure}

Following point 1, a first systematic reduction of the Hilbert space could be obtained by
directly working in the singlet sector $S=0$. Unfortunately, ED are not adapted to an explicit
implementation of the SU(2) symmetry of the Heisenberg Hamiltonian because, from a numerical point
of view, eigenvectors of the total spin ${\bf S}^2$ turn to be very complex objects for large systems. In practice,
the Hilbert space used in ED is a set of $S^z=0$ eigenvectors. The expected benefit
of such a reduction would be~\cite{Sizes} of order $\sim 1/N$.

In fact, a natural framework for fixed ${\bf S}^2$ states has been 
developed years ago\cite{Singlets}. Indeed, the whole
singlet subspace can be generated using arbitrary range coverings of the 
lattice with VB states (see figure \ref{fig:srrvb} {\bf (a)})~:
\begin{equation}
\label{eq:rvb}
\vert \psi \rangle = \prod_{(i,j)} [i,j],
\end{equation}
where $[i,j] = (1/\sqrt{2}) (\vert \uparrow_i \downarrow_j \rangle - 
\vert \downarrow_i \uparrow_j \rangle)$.
However, the practical relevance of these states is very limited because the 
number of dimer coverings
for the complete graph is $ N! / (2^{N/2} (N/2)!) \sim (N/e)^{N/2}$ which is 
much larger than the size
of the singlet subspace. As a direct consequence, this family of states is 
overcomplete. Furthermore
it is certainly not specifically adapted to the description of non-magnetic 
(quantum disordered) 
phases since any kind of singlet state, including the finite-size N\'eel state,
could be constructed by an appropriate linear combination of arbitrary range 
VB states.

Let us now examine point 2. A simple way to reduce the number of coverings 
while keeping only short range correlations
is to restrict the range of the dimers to short-range, for example nearest 
neighbor valence bond states (NNVB) (see figure \ref{fig:srrvb} {\bf (b)}). 
A general
solution to the question of enumerating these states has been given by 
Fisher\cite{Fisher}. It is exponential $k \alpha^N$ for large $N$,
with $\alpha \approx 1.34 $ (square lattice),  $\alpha \approx 1.53 $ 
(triangular lattice) and $\alpha \approx 1.26$ (kagome lattice).
As expected, these numbers are much smaller than the total number of 
singlets thus providing the desired selection inside the singlet sector.
Nevertheless, two important questions deserve attention : (i) Are these states linearly independent ?
(ii) Which class of singlet states can be obtained by linear combinations of NNVB states ? The first question has not been addressed 
analytically but numerical calculations\cite{Unpublished} show that, unless for very small systems on the triangular lattice, these states
are linearly independent for the square, triangular and kagome lattices. 
Concerning the second question, it is clear that any state involving only short range spin-spin correlations, from VBC to SL,
can be captured by SRVB states. On the contrary, Liang {\it et al.} showed\cite{LDA} that magnetic long range order cannot
be obtained from linear combinations of such configurations.

As a partial conclusion, selecting a subset of SRVB states in the singlet space provides a convenient framework to study the low-energy singlet sector
of highly frustrated antiferromagnets. If the physics of a given problem can be captured in this restricted basis, this kind of approach not only 
makes larger systems accessible to computation but also gives some insights about the nature of the GS ruling out any magnetic long range order.
For technical details and illustrations of the method the reader can refer 
to previous publications\cite{KagomeRVB,checkerboard,StaticKagomeImpurities}.
Nevertheless, let us recall one of the most salient characteristic of the 
calculation : one crucial property of SRVB states is their non-orthogonality 
(see appendix).
At a numerical level, the problem of diagonalizing the Hamiltonian is then shifted to the so called generalized eigenvalue problem (GEP): 
\begin{equation}
\label{eq:generalized}
\det \left ( {\cal H} - E {\cal O} \right )=0,
\end{equation}
where $\cal O$ denotes the overlap matrix. The GEP, especially when ${\cal H}$ and ${\cal O}$ are non-sparse matrices, cannot
be efficiently solved iteratively. A rather time consuming complete diagonalization has to be performed, which makes use of
spatial symmetries necessary for large clusters. 

Finally, let us remark that the GS computed with this method can be seen as the best variational approximation of the exact GS using the restricted NNVB subset of states.
However, even for a magnetically disordered exact ground state, the wave function almost certainly involves still finite range but more than only nearest neighbor VB states. As a consequence, this approach is not designed to provide the state-of-the-art variational approximation of the exact GS, but rather
to capture in a small subset of physically suggestive states, the main part of the absolute ground state wave function neglecting finite range correlations refinements whose sole effect would be to slightly renormalize energies. In this respect,  solving (\ref{eq:generalized}) is exactly equivalent to diagonalize a sophisticated effective Hamiltonian, namely the exact projection of the Heisenberg Hamiltonian on the chosen SRVB subspace.

\section{SRVB region}

One of the main drawbacks of the SRVB method is its lack of built-in 
control~: solving (\ref{eq:generalized}) is always possible even if the 
selected SRVB subspace is irrelevant to describe the low-energy sector of 
${\cal H}$. It is therefore necessary to make systematic comparisons between 
SRVB results and exact ones.

\begin{figure}
\centerline{\includegraphics*[angle=0,width=0.9\linewidth]{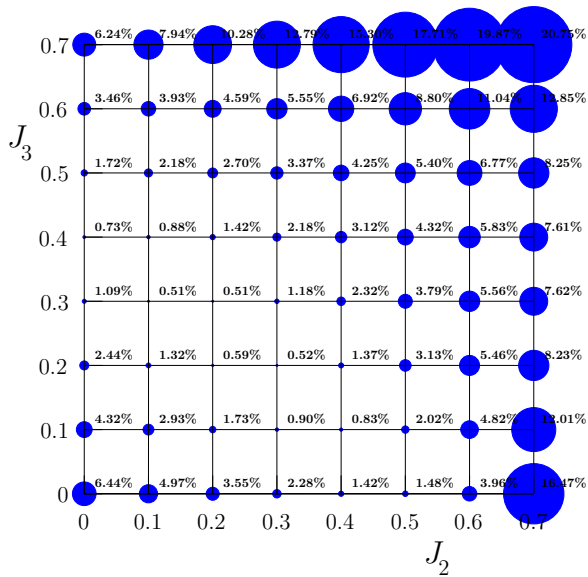}}
  \caption{\label{fig:edrvb}
Systematic comparison of ED and NNVB ground state energy for $N=32$. The radius of the circles is proportional to the
NNVB ground state accuracy $(E_0^{\text{NNVB}}-E_0^{\text{ED}})/E_0^{\text{ED}}$. Typical values of the energies are given in table \ref{tab:energies}.}
\end{figure}

To do so, let us consider an intermediate size cluster, namely $N=32$, and compute the GS energy both by ED and NNVB diagonalizations, respectively 
$E_0^{\text{ED}}$ and $E_0^{\text{NNVB}}$. The accuracy and thus the validity of NNVB approach can be tested by a measurement  of the parameter $(E_0^{\text{NNVB}}-E_0^{\text{ED}})/E_0^{\text{ED}}$. On figure \ref{fig:edrvb}, this quantity is plotted as a function of $J_2 / J_1$ and $J_3/ J_1$.

As expected, the NNVB ground state fails to approximate the exact one in the regions of the phase diagrams known to be magnetically ordered : $(J_2 \ll J_1, J_3 \ll J_1)$, $(J_2 \gg J_1, J_3 \ll J_1)$ or $(J_3 \gg J_1, J_2 \ll J_1)$. On the opposite, in the highly frustrated regime, an extended region of the phase diagram emerges around $(J_2 + J_3)/J_1 \sim  1/2$ where $(E_0^{\text{NNVB}}-E_0^{\text{ED}})/E_0^{\text{ED}}$ is smaller than $1.5 \%$ and as small as $0.5\%$ (see figure \ref{fig:edrvb} and table \ref{tab:energies}).
\begin{table}[htbp]
    \newcommand\TT{\rule{0pt}{2.6ex}}
    \newcommand\BB{\rule[-1.2ex]{0pt}{0pt}}
    \centering\footnotesize
        \begin{tabular}{|c|c|c||c|c|c|}     
        \hline \hline
         $( J_2 , J_3)$ \TT \BB & $E_0^{\text{ED}}$ & $E_0^{\text{NNVB}}$ & $( J_2 , J_3 ) \TT \BB$ & $E_0^{\text{ED}}$ & $E_0^{\text{NNVB}}$ \\ \hline
        $(0.0,0.3)$ \TT \BB & -18.71704 & -18.51215 & $(0.2,0.4)$ \TT \BB & -16.66878 & -16.43163 \\ \hline
        $(0.0,0.4)$ \TT \BB & -18.12399 & -17.99224 & $(0.3,0.1)$ \TT \BB & -17.12863 & -16.97424 \\ \hline
        $(0.0,0.5)$ \TT \BB & -17.92509 & -17.61700 & $(0.3,0.2)$ \TT \BB & -16.50461 & -16.41946 \\ \hline
        $(0.1,0.2)$ \TT \BB & -18.43435 & -18.19099 & $(0.3,0.3)$ \TT \BB & -16.17630 & -15.98600 \\ \hline
        $(0.1,0.3)$ \TT \BB & -17.72089 & -17.63119 & $(0.4,0.0)$ \TT \BB & -16.90813 & -16.66731 \\ \hline
        $(0.1,0.4)$ \TT \BB & -17.33094 & -17.17892 & $(0.4,0.1)$ \TT \BB & -16.21783 & -16.08331 \\ \hline
        $(0.2,0.2)$ \TT \BB & -17.39604 & -17.29400 & $(0.4,0.2)$ \TT \BB & -15.80152 & -15.58522 \\ \hline
        $(0.2,0.3)$ \TT \BB & -16.86835 & -16.78183 & $(0.5,0.0)$ \TT \BB & -16.00307 & -15.76633 \\ \hline \hline 
        \end{tabular}
    \caption{ED and NNVB ground state energy for $N=32$ as a function of $(J_2,J_3)$ (units of $J_1$).}
    \label{tab:energies}
\end{table}

Before going any further in the analysis, it is important to have in mind the order of magnitude of the NNVB truncation of the Hilbert space. For such a system size, the dimension of the GS representation (${\bf k} = (0,0)$, {\it s-wave}) is 
$1184480$. This has to be compared to the number of NNVB configurations in 
the same representation which is only 182. The reduction factor is thus $\sim 10^4$. 

\begin{figure}
\centerline{\includegraphics*[angle=0,width=0.9\linewidth]{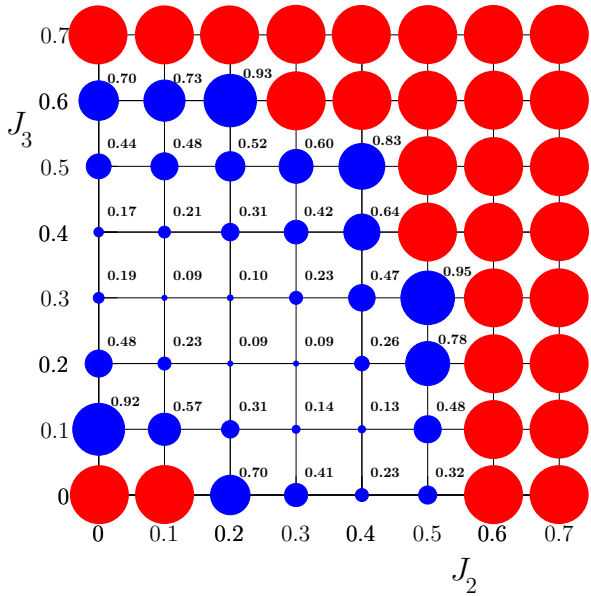}}
  \caption{\label{fig:edrvb2} (Color online) Upper bound for the ``missing weight''
$1-\vert  \langle \psi_0 \vert \psi_0^{\text{NNVB}}\rangle \vert^2$. The radius of the blue circles is proportional to the upper bound given in equation (\ref{eq:upper}). Values greater than one being irrelevant are represented as unit radius red circles.}
\end{figure}

Considering both the accuracy of $E_0^{\text{NNVB}}$ {\it and} the rather 
drastic reduction of the singlet space we can conclude to the existence of an extended region in the phase diagram, around $(J_2+J_3)/J_1 \sim 1/2$, where 
the exact GS can be described with only NNVB states. Nevertheless, in order to investigate the precise nature of the ground state using this wave function, 
it is important to go beyond this energetic criterion. A direct evaluation
of the overlap between the exact ground state and the NNVB variational wave 
function $  \langle \psi_0 \vert \psi_0^{\text{NNVB}}\rangle$ cannot be done 
easily, but it is straightforward to compute an upper bound for the so-called 
``missing weight''
$1-\vert \langle \psi_0 
\vert \psi_0^{\text{NNVB}}\rangle \vert^2$ which, crudely, quantifies the ``accuracy'' of the
wavefunction w.r.t. the exact GS. A formal normalized expansion of 
$\vert \psi_0^{\text{NNVB}}\rangle = \sum_i \alpha_i \vert \psi_i \rangle$ 
on the exact eigenstates leads to the expression of 
$E_0^{\text{NNVB}}=\sum_i \vert \alpha_i
\vert^2 E_i$ as a function of the exact eigenenergies $E_i$. Since $E_i \geq E_1$ for $i>1$ one obtains,
\begin{equation}
\label{eq:upper}
1-\vert \langle \psi_0 \vert \psi_0^{\text{NNVB}}\rangle \vert^2 \leq \min{ \left ( \frac{E_0^{\text{NNVB}}-E_0}{E_1-E_0},1 \right )}.
\end{equation}
This quantity is represented on figure \ref{fig:edrvb2} as a function of $J_2/J_1$ and $J_3/J_1$. 
Despite the fact that this upper bound is far from being optimal since $E_1$ is only a crude lower bond 
for highly excited states, the same region of the phase diagram (as the one determined previously on a purely energetic criterion) emerges where  
$\vert \langle \psi_0 \vert \psi_0^{\text{NNVB}}\rangle\vert$ is at least $90\%$ in the worst 
case and  up to $95\%$ in the best case. 

This picture clearly confirms that around $(J_2 + J_3)/J_1 \sim 1/2$ the essential part of the GS wave 
function can be captured using only few SRVB states, namely NNVB configurations. As mentioned in 
the previous section, there is no doubt that this accuracy could be systematically improved by 
dressing $\vert \psi_0^{\text{NNVB}}\rangle$ with some longer (but still finite) range VB configurations 
(eg next nearest neighbor VB configurations). Although including 
such additional configurations are expected to
lower even further the variational 
energy, this would be no more than refinements and we believe that the 
approach here already fully captures the physical 
picture of a SRVB ground state.

\section{Dimer-dimer correlations and structure factors}

The next important question is now to investigate the nature, VBC or SL, of the SRVB ground state in this region. To address this question
we used the SRVB method to compute the dimer-dimer correlation function~:
\begin{equation}
\label{eq:dimerdimer}
C_{ijkl} = \langle ({\bf S}_i .{\bf S}_j )({\bf S}_k .{\bf S}_l)\rangle - (\langle {\bf S}_i .{\bf S}_j \rangle)^2.
\end{equation}
The SRVB method allows a systematic computation of (\ref{eq:dimerdimer}) on the extended SRVB phase for cluster sizes ranging from $N=20$ to $N=40$.

\begin{figure}
\centerline{\includegraphics*[angle=0,width=0.9\linewidth]{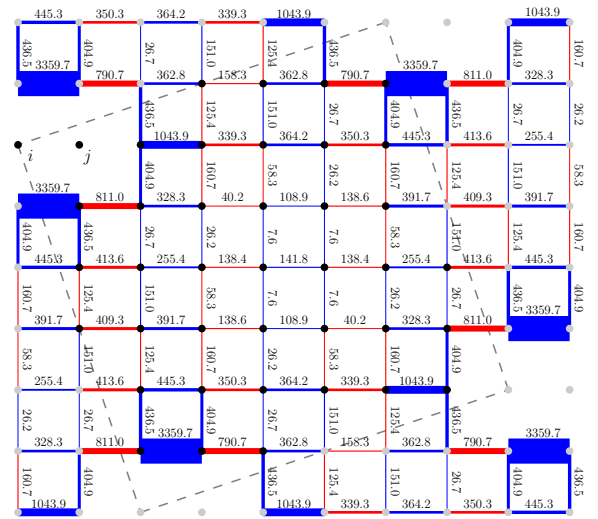}}
  \caption{\label{fig:dimerJ2}
(Color online) Dimer-dimer correlation function for a 40 site cluster with periodic boundary conditions with $J_2/J_1=1/2$ and $J_3=0$. The dashed line delimits the cluster, $(i,j)$ is the reference bond, and the width of the solid bonds $(k,l)$ are proportional to the absolute values of $C_{ijkl}$. The blue (resp. red) bonds denote positive (resp. negative) correlations. Numbers correspond to $\vert 10^4 C_{ijkl} \vert$.}
\end{figure}

\begin{figure}
\centerline{\includegraphics*[angle=0,width=0.9\linewidth]{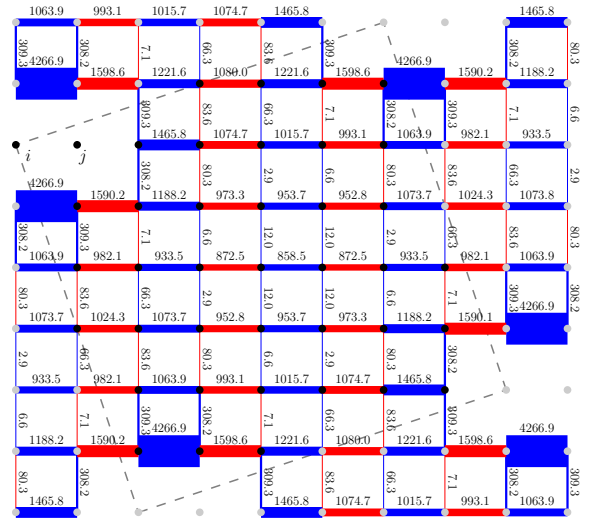}}
  \caption{\label{fig:dimerJ3}
(Color online) Same as figure \ref{fig:dimerJ2} for $J_3/J_1=1/2$ and $J_2=0$.}
\end{figure}

{\it Real space picture.} Figures \ref{fig:dimerJ2} and \ref{fig:dimerJ3} are snapshots of the results for $N=40$ respectively for the pure $J_1-J_2$ model at $J_2/J_1=1/2$ and the pure  $J_1-J_3$ model at $J_3/J_1=1/2$. Both systems exhibit, for bonds parallel to the reference bond $(i,j)$, a clear alternating pattern of correlated and anticorrelated rows. 
Moreover, around the maximal distance from the reference bond, the values of the parallel correlations are almost constant. As a consequence both figures \ref{fig:dimerJ2} and \ref{fig:dimerJ3} suggest a translational symmetry breaking VBC phase with a stronger signal in the latter case.

As suggestive as this kind of picture may be, two important question have to be addressed : (i) To what kind of VBC phase figures \ref{fig:dimerJ2} and \ref{fig:dimerJ3} correspond ? (ii) Is this suggested long range order robust when $N \rightarrow \infty$ ?

Even if at first sight these real space pictures naively suggest a columnar arrangement, a plaquette VBC order 
cannot be ruled out. In order to investigate the nature of the VBC ground state, we introduce  3 trial wave functions $\psi_c$, $\psi_s$ and $\psi_d$
respectively referring to a columnar, {\it s-wave} plaquette and {\it d-wave} plaquette state (see Appendix). These wave functions are designed
to have the same symmetry as the finite size ground state, namely ${\bf k}=(0,0)$ {\it s-wave}, in order to allow direct comparisons with the numerical results.

The computation of the dimer-dimer correlations in these wavefunctions is presented in details in the 
Appendix and the results are summarized in table \ref{tab:correlationsresult}.
First, a comparison of our previous numerical results with those of table \ref{tab:correlationsresult}
shows that the {\it d-wave} plaquette scenario is very unlikely. 
Furthermore, the results of the Appendix suggest that the key criterion to discriminate between a 
pure columnar and a pure {\it s-wave} plaquette VBC, on the basis of dimer-dimer correlations, 
is the ratio between (i) perpendicular bond correlations (with respect to the reference bond) 
and (ii) parallel bond correlations in odd columns (defining the reference bond column as even). 
In the first case it is expected to be equal to 1 while it should vanish in the latter case.  

For the data shown in figures \ref{fig:dimerJ2} and \ref{fig:dimerJ3}, if one considers the most distant
bonds from the reference one, the typical value of this ratio is of order $1/20$ and $1/100$ respectively. This strongly supports a {\it s-wave} plaquette scenario for $J_3=J_1/2$ and $J_2=0$ while the situation appears more involved for $J_2=J_1/2$ and $J_3=0$ where the ratio is still very small but for a much weaker overall long range correlation signal.

\begin{figure}
\centerline{\includegraphics*[angle=0,width=\linewidth]{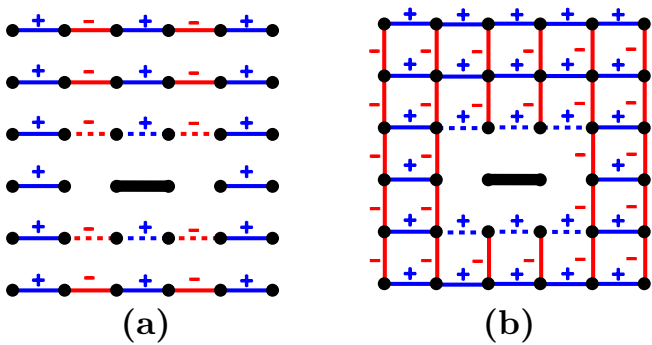}}
  \caption{\label{fig:structure_factors}
(Color online) Phase factors $\varepsilon_\lambda (k,l)$ for structure factors : {\bf (a)} ``VBC" {\bf (b)} ``col".
Dashed bonds are either included or excluded form the definition (\ref{eq:structure_factors}) in the fitting procedure
in order to test the sensitivity of the extrapolation scheme to irrelevant short range contributions.  
Note that the $(k,l)$ bonds nearest neighbors
to the reference one (central black solid bond)
are always omitted in the sum defining $S_{\lambda}$.}
\end{figure}

{\it Finite size analysis and structure factors}. It is crucial to study the robustness of this picture with the system size. A convenient way to investigate the thermodynamic limit is to introduce spatially integrated quantities such as dimer
structure factors and perform finite size scaling.  The essential difference between columnar and {\it s-wave} plaquette orders is the breakdown of rotational symmetry. Following Ref. \onlinecite{j1j2}, it is possible 
to build
two structure factors $S_{\text{VBC}}$ and $S_{\text{col}}$ with the following properties : 
\begin{itemize}
\item $S_{\text{VBC}}$ diverges at thermodynamic limit both in columnar and plaquette states,
\item $S_{\text{col}}$ diverges at thermodynamic limit only in a columnar state.
\end{itemize}
To achieve this, the form factors introduced in $S_{\text{VBC}}$ and $S_{\text{col}}$ have to reflect the patterns 
of table \ref{tab:correlationsresult}.
It is easy to verify that appropriate structure factors can be defined e.g. as~:
\begin{equation}
\label{eq:structure_factors}
S_{\lambda} = \sum_{(k,l)} \varepsilon_{\lambda} (k,l) C_{ijkl}, 
\end{equation}
where $\lambda$ stands for either ``VBC'' or ``col'' and
the corresponding form factors $\epsilon_{\lambda} (k,l)$ are defined according to figure 
\ref{fig:structure_factors}.

In an ordered phase $S_{\lambda}$ is extensive so that $S_{\lambda} / N_b$, where $N_b$ denotes the
number of bonds involved in (\ref{eq:structure_factors}), 
is expected to scale like $C^{\infty}_{\lambda}+A/N$ with $C^{\infty}_{\lambda}$ being the 
square of the {\em bond} order parameter
in the thermodynamic limit. The divergence (resp. finite value) of $S_{\lambda}$ is thus signaled 
by a finite (resp. vanishing) $C^{\infty}_{\lambda}$.

\begin{figure}
\centerline{\includegraphics*[angle=0,width=\linewidth]{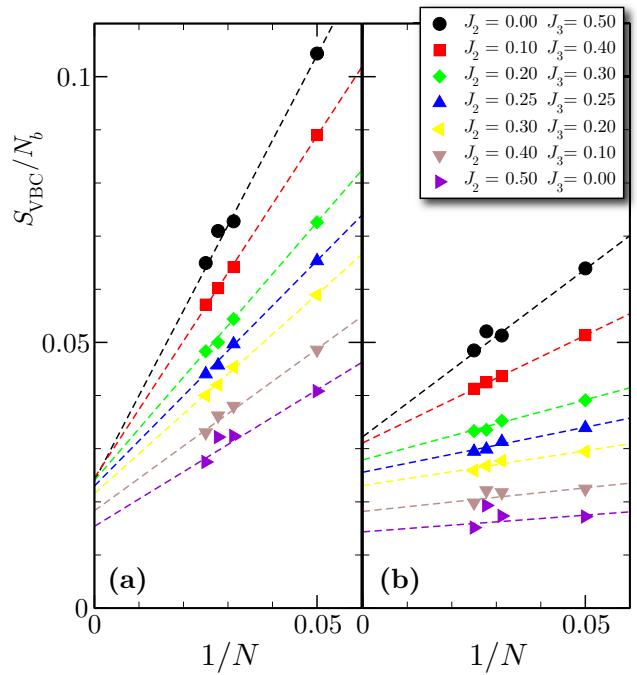}}
  \caption{\label{fig:chi_vbc}
(Color online) {\it Left panel} {\bf (a) :} $S_{\text{VBC}}/N_b$ as a function of $1 / N$ along the $J_2 + J_3 = J_1 / 2$ line. 
Note that the $N=36$ data is excluded from the linear fits represented as dashed lines. {\it Right panel} {\bf (b) :} Same as left panel
with a modified definition of $S_{\text{VBC}}$ in which very short range contributions are excluded (dashed bonds in figure \ref{fig:structure_factors}).}
\end{figure}

\begin{figure}
\centerline{\includegraphics*[angle=0,width=\linewidth]{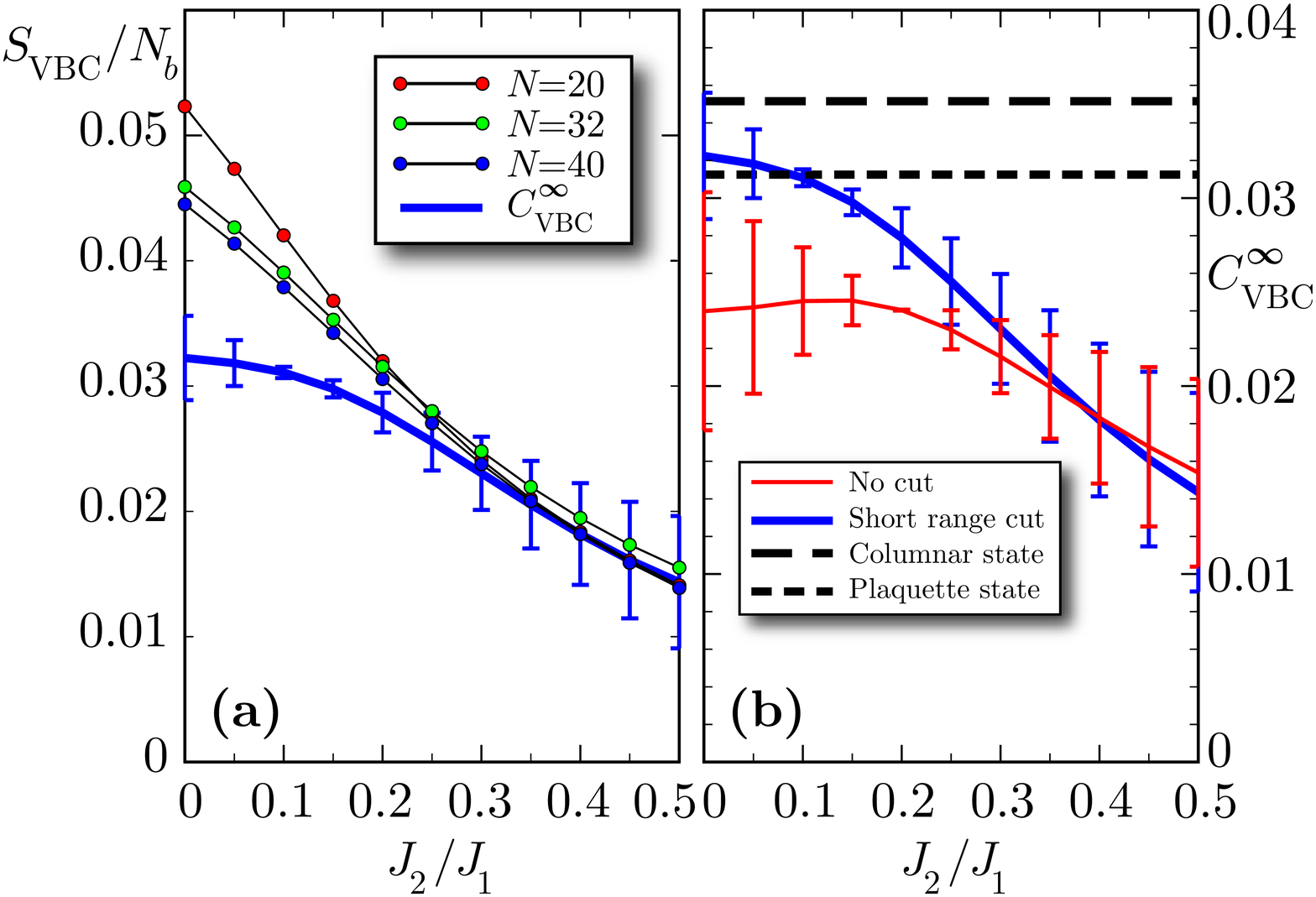}}
  \caption{\label{fig:cut}
(Color online) {\it Left panel} {\bf (a)} : Extrapolation $C^{\infty}_{\text{VBC}}$ of $S_{\text{VBC}}/N_b$ as a function of $J_2/J_1$ along the $J_2+J_3=J_1/2$ line (Thick solid line with error bars). Finite size $S_{\text{VBC}}/N_b$ data for $N=20,32$ and $40$ are represented as thin lines and circles. The error bars reflect the quality of the $1/N$ fit presented on figure \ref{fig:chi_vbc} {\bf (b)}. {\it Right panel} {\bf (b)} : Influence of short range contributions to the extrapolated VBC structure factor 
$C^{\infty}_{\text{VBC}}$ along the line $(J_3+J_2)=1/2$ as a function of $J_2/J_1$ and comparison with expected values for columnar and {\it s-wave} plaquette states. Thin (resp. thick) line with error bars labelled ``Not cut'' (resp. ``Short range cut'') corresponds to the results of the fits of $S_{\text{VBC}}/N_b$ including all range contributions (resp. excluding short range contributions) represented on figure \ref{fig:chi_vbc} {\bf (a)} (resp. {\bf (b)}). Thick dashed lines are the expectations values of the structure factors
$C^{\infty}_{\text{VBC}}$ at thermodynamic limit for the pure columnar (short dashed line) state $\psi_c$ and the pure {\it s-wave} plaquette state $\psi_s$.}
\end{figure}

We performed this type of scaling on $S_{\text{VBC}}/N_b$ for $N=20,32,36$ and 40 
along the line $(J_2+J_3)/J_1 = 1/2$. As shown in 
figure \ref{fig:chi_vbc} {\bf (a)}, the quality of a $1/N$ extrapolation is greatly affected by the 
$N=36$ data. This point is due to the peculiar shape of this cluster whose periodic boundary conditions induce short loops that have the tendency to 
overestimate the influence of the reference bond and thus
$S_\lambda$. We therefore excluded this set of data in the analysis depicted on figure \ref{fig:chi_vbc} {\bf (a)}. 
Along the whole $(J_2+J_3)/J_1 = 1/2$ line, the fit reveals a non-vanishing extrapolated $C^{\infty}_{\text{VBC}}$ and a standard evaluation of errors
bars on the extrapolated values is presented on figure \ref{fig:cut} {\bf (b)} (thin line labelled ``No cut").

>From a technical point of view it is fair to evaluate, in the extrapolation scheme, the influence of 
the strong contributions to the structure factor coming from the short range part of the dimer-dimer 
correlations (see figures \ref{fig:dimerJ2} and \ref{fig:dimerJ3}). There are at least two reasons
to discuss this aspect~: (i) The short range part of the data is irrelevant
at large distance and therefore, a non negligible contribution to the thermodynamic extrapolation would indeed be problematic (ii) As shown in the Appendix,
a substantial enhancement of the short range dimer-dimer correlations is expected to occur in plaquette states (see table \ref{tab:correlationsresult}).

The sensitivity of the fit to the (irrelevant) short range correlations
can be tested by systematically removing from the sum defining the structure factor the contribution
of the neighboring bonds of the reference one (see dashed bonds in figure \ref{fig:structure_factors}). 
As shown in figures \ref{fig:chi_vbc} and \ref{fig:cut} {\bf (b)}, when
$J_2 / J_1\rightarrow 1/2$ the extrapolated values of $S_{\text{VBC}}$ is 
insensitive to the short 
range correlations, while in the crystalline phase, the procedure of removing the short-range part
of the data has a systematic tendency to enhance the VBC order parameter and to lower the error bars thus improving the confidence of the
extrapolated value.
This fact convincingly establishes that the underlying GS has a VBC long order for $J_2 / J_1 \leq 0.2 - 0.3$ but also 
gives some further indication: very short range dimer-dimer correlations in the GS are responsible for a slight perturbation of the extrapolation which is compatible with the local enhancement of $C_{ijkl}$ observed in the trial plaquette state $\psi_s$ when $(k,l)$ is lying next to $(i,j)$ (see table \ref{tab:correlationsresult}). From a technical point of view, in order to exclude this kind of short range effect, we exclude for further analysis the short distance contribution to the definition (\ref{eq:structure_factors}) of $S_\lambda$.

A careful inspection of figure \ref{fig:chi_vbc} reveals two regimes of 
parameters for $J_2 / J_1$~: below $\sim 0.2$ the opening of the errors bar is due to to a {\em convex} deviation from a perfect linear behavior, while it is {\em concave} above $J_2 / J_1 \sim 0.3$. As a consequence, the extrapolation scheme respectively underestimates and overestimates $S_{\text{VBC}}$. This  confirms that the crystalline order is indeed robust for 
$J_2 / J_1 \leq 0.2 - 0.3$. Moreover 
the extrapolated value for $J_2/J_1 = 0$ (and $J_3/J_1 = 1/2$) 
is $0.032 \pm 0.003$ which compares very well with the expected 
values of 
the pure columnar or plaquette crystalline states which respectively equal to $9/256 \sim 0.035$ and
$1/32 \sim 0.031$ (see table \ref{tab:correlationsresult} in the Appendix and figure \ref{fig:cut} {\bf (b)}). 

In contrast, due to large error bars and the slight concavity of the $1/N$ extrapolation, a vanishing $C^{\infty}_{\text VBC}$ cannot be ruled out from our data for  $J_2 / J_1$ larger than 0.3 and therefore the existence of a crystalline long range order for $J_3=0$ is not proven by the present calculation.

\begin{figure}
\centerline{\includegraphics*[angle=0,width=\linewidth]{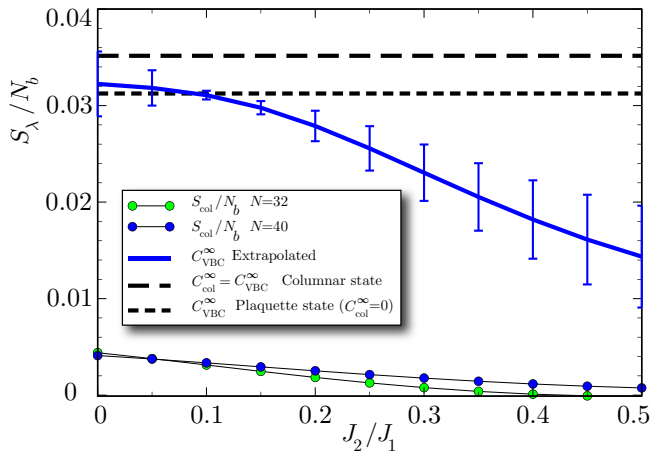}}
  \caption{\label{fig:col_plaq}
  Comparison between $S_{\text{VBC}}/N_b$  and $S_{\text{col}}/N_b$ as a function of $J_2/J_1$ along the $J_2+J_3 = J_1/2$ line. Thin lines with circles~: finite size data for  $\chi_{\text{col}}/N_b$ for $N=32$ and $N=40$. Thick dashed lines are the expectations values of the structure factors
$C^{\infty}_{\text{VBC}}$ and $C^{\infty}_{\text{col}}$ at thermodynamic limit for the pure columnar (short dashed line) state $\psi_c$ and the pure {\it s-wave} plaquette state $\psi_s$. Note that $C^{\infty}_{\text{col}}=0$ in the {\it s-wave} plaquette state. The thick line with errors bars is the same
as in figure~\ref{fig:cut}.}
\end{figure}

Let us now turn to $S_{\text{col}}$. The size dependence of 
$S_{\text{col}}/N_b$  does not allow a confident extrapolation to obtain 
$C^{\infty}_{\text{col}}$ with enough accuracy. 
Nevertheless, for all clusters $S_{\text{col}}$ is always a very small fraction of $S_{\text{VBC}}$ as
can be seen by comparing figures \ref{fig:cut} {\bf (a)} and \ref{fig:col_plaq} for $N=32$ and $N=40$. 
Typically the ratio $S_{\text{col}}/S_{\text{VBC}}$ is of order $1/20$ 
for $J_2 / J_1 = 0$ and $1/15$ for $J_3 / J_1 = 0$. The expected values of this ratio for the pure 
columnar and {\it s-wave} plaquette state (see table \ref{tab:correlationsresult}) are respectively 1 and 0. 

We cannot draw definitive conclusions from our data in the regime where $J_2 / J_1\sim 1/2$ 
and $J_3\rightarrow 0$ 
since our scaling does not exclude a scenario where $C^{\infty}_{\text{VBC}}$ 
and $C^{\infty}_{\text{col}}$ would vanish. In contrast, on the $(J_3+J_2)/J_1=1/2$ line for small $J_2$ and up to $J_2/J_1 \sim 0.3$,
the fact that $S_{\text{col}}$ is much weaker than $S_{\text{VBC}}$ 
is very much in favor of the {\it s-wave} plaquette scenario with an 
{\it absence of rotational symmetry breaking} and seems to rule out
a simple long range columnar order for which $S_{\text{col}}\approx S_{\text{VBC}}$ in the thermodynamic limit. Note that a small spatial anisotropy of the plaquette phase is still possible. This scenario where the vertical and horizontal bond amplitudes within the resonating plaquettes
are slightly different would indeed lead to a small value of the columnar structure factor in the thermodynamic limit and a 
GS degeneracy of 8 (instead of 4)~\cite{note-ref}.


\section{Plaquette-plaquette correlations}
\begin{figure*}
 \centerline{\includegraphics[width=\linewidth]{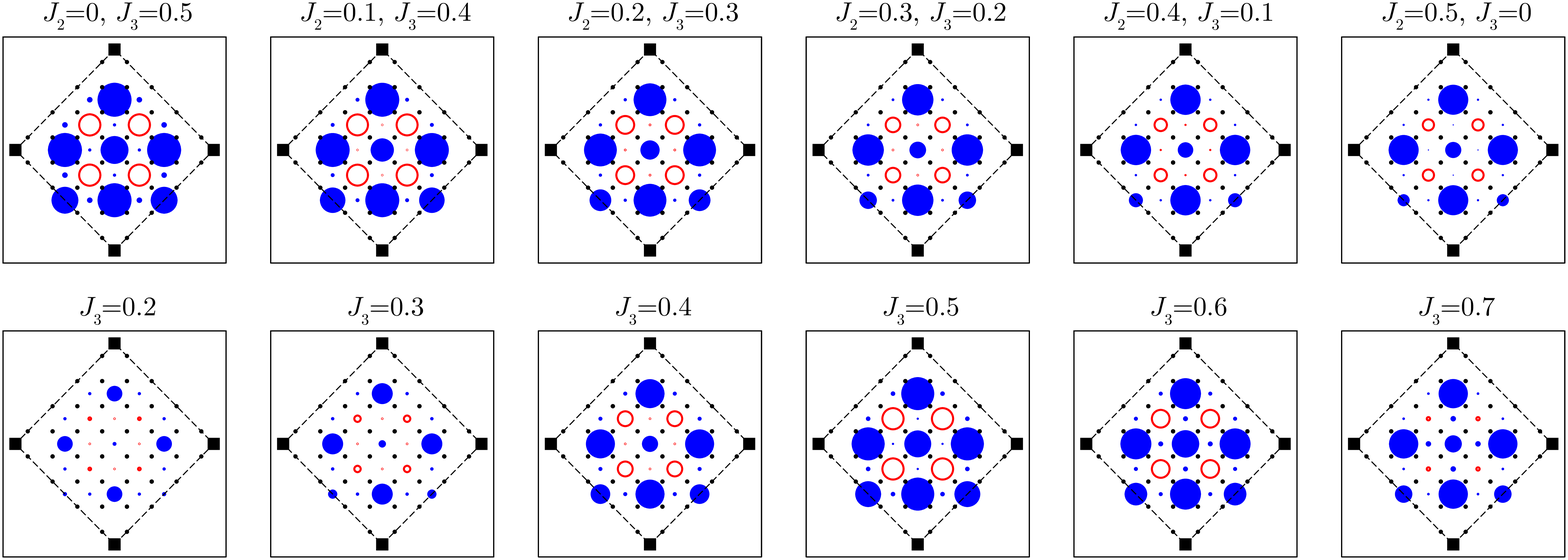}}
  \caption{\label{fig:plaquetteED}
  Plaquette correlation function $\mathcal{C}_\mathrm{Plaquettes}(p,q)$ [Eqn.~(\protect{\ref{eqn:plaquette}})], obtained by exact diagonalization on a $N=32$ sample.
  The black squares denote the reference plaquette, filled blue circles correspond to positive values and empty red circles denote
  negative values. The tiny black circles mark the locations of the sites. 
  The correlations are in excellent agreement with the qualitative expectations of a plaquette phase, especially
  around $J_2=0, J_3/J_1=1/2$.
  {\it Upper row:} Correlations along the line $(J_2+J_3)/J_1=1/2$.
  {\it Lower row:} Correlations for $J_3/J_1 \in [0.2,0.7]$, while $J_2$ is set to zero.
  }
\end{figure*}
A careful analysis of the difference in dimer-dimer correlations in a columnar dimer versus an {\it s-wave} plaquette ordered 
singlet state performed in the previous section yielded strong support for a plaquette phase. 
In order to directly image the plaquettes in real space we calculate the following 8-spin correlation function using ED:
\begin{eqnarray}
\mathcal{C}_\mathrm{Plaquettes}(p,q)&=& \langle Q_p  Q_q \rangle - \langle Q_p \rangle^2 \label{eqn:plaquette} \\
Q_p&\equiv&P ^{\phantom{-1}}_{\Box ,p} + P^{-1}_{\Box  ,p} \nonumber
\end{eqnarray}
where $p$ and $q$ denote two different plaquettes and $P^{\phantom{-1}}_{\Box ,p}$ denotes the cyclic exchange operator of
the four spins on a given plaquette. This correlation function has also been used in a recent study of plaquette order in the
checkerboard antiferromagnet~\cite{checkerboard}. 
If we want to discriminate between a columnar dimer state and a plaquette ordered state in the following, it is useful to 
note that in a columnar dimer state one has {\em two} distinct expectation values of $ \langle Q_p \rangle$ 
(either covering two singlet bonds or none), whereas in a plaquette ordered state we expect {\em three} distinct expectation 
values (on a singlet plaquette, between two adjacent singlet plaquettes, or sharing the corners of four distinct singlet plaquettes). 
This number is expected to translate into the number of different values in the correlation function Eqn.~(\ref{eqn:plaquette}).

We present the results obtained by ED on a $N=32$ sample in Fig.~\ref{fig:plaquetteED}, both along a line with 
$(J_2+J_3)/J_1=1/2$ (upper row) and along the pure $J_3$ line (lower row).
In the cases where strong correlations are seen, we basically detect {\em three} different types
correlation function values, in agreement with the expectations of the plaquette phase, as pointed out above. 
Furthermore the spatial structure coincides with the plaquette picture, i.e both the positively and the negatively 
correlated plaquettes form a distinct $2\times 2$ superlattice, shifted by the vector (1,1) with respect to each other.
The evolution of the correlations as a function of $J_2$ and $J_3$ shows that the strength of the correlations both 
decreases as one moves away from the point $J_2=0, J_3/J_1=1/2$ either along the pure $J_3$ line or along the line 
with fixed $(J_2+J_3)/J_1=1/2$, in agreement with the results of the preceding section based on dimer-dimer correlations.
Interestingly the correlations at the much debated point $J_2/J_1=1/2, J_3=0$ are rather weak, but still carry some 
remnants of the plaquette phase, at least for this $N=32$ sample.

\section{Discussion and Conclusions}

An extensive numerical study of the Heisenberg $J_1 - J_2 - J_3$ antiferromagnet using both exact 
diagonalizations
and a short range valence bond method shows that, in the most frustrated part of the phase diagram 
(around $J_2 + J_3  \sim J_1 / 2$),
the ground state can be captured using only nearest neighbor valence bond coverings of the square 
lattice. 
The emergence at low energy of short range valence bond singlet physics for these parameters and 
thus the breakdown of magnetic long range order
is a direct consequence of the strong frustration of the model.
Moreover, we characterize the ground state by an analysis of dimer-dimer correlations, dimer 
structure factors and plaquette-plaquette correlations
and show numerical evidences for an extended valence bond crystal phase around $J_2 + J_3 \simeq J_1 /2$ 
and $J_2 \leq J_3$ where the ground
state is a {\it s-wave} plaquette state only breaking translational symmetry. As a consequence, the $J_1-J_3$ model
provides an example of frustration-driven N\'eel to VBC quantum phase transition. Note
that the SRVB framework can be readily extended to include singlet pairs beyond nearest neighbors. However, we believe
that this will modify only slightly the results in the maximally frustrated region where the magnetic correlation length is very small. Such
an approach could nevertheless be useful to investigate properties close to the critical point where the spin correlation length is expected to grow. In that respect such a transition can be probed by introducing static (non magnetic) impurities~\cite{j1j2j3_doped}. Again, our framework could be extended to that case~\cite{StaticKagomeImpurities}.

For $J_3 \leq J_2$, including the much debated frustrated phase of the pure $J_1-J_2$ model,
the NNVB description of the ground state remains relatively robust. While our results are not
able to resolve the controversy around $J_2/J_1 \approx 1/2$, the inclusion
of an additional $J_3$ coupling allows us to put this region into a broader
perspective. We show that an antiferromagnetic $J_3$ is useful in pushing
the magnetically ordered phases further apart, therefore leaving more
room for the disordered phases, and enabling us to reveal a robust
plaquette singlet ordered phase. On the contrary, a ferromagnetic $J_3$
interaction will probably lead to a direct first order transition between
the $(\pi,\pi)$ and the $(\pi,0)$ N\'eel order phases as function of $J_2$, similar
to the classical analysis and numerical results on the related bcc
lattice \cite{bccj1j2}.
The closeness of the magnetically order phases and the related phase
transitions are probably responsible for the enormous difficulty
in settling the controversy on the nature of the magnetically disordered phase(s)
of the pure $J_1-J_2$ model on the square lattice.


\acknowledgements

We acknowledge useful discussions with F.~Becca, S.~Capponi and F.~Assaad. We thank IDRIS (Orsay, France)  and CSCS (Manno, Switzerland)
for allocation of CPU-time on the IBM-SP4 supercomputers.
This work was supported by the Swiss National Fund, MaNEP and the Agence Nationale de la Recherche (France).

\appendix*
\section{VB states properties and dimer-dimer correlations for some relevant VBC states}

In this appendix, we recall some basic overlap properties of VB states and compute dimer-dimer correlations expectation 
values for columnar and plaquette states.

\begin{figure}
\centerline{\includegraphics*[angle=0,width=\linewidth]{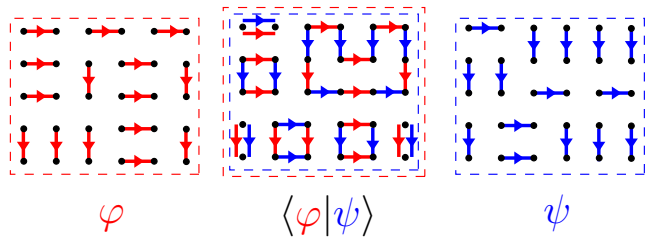}}
  \caption{\label{fig:overlap}
Overlaps between two VB states.}
\end{figure}

{\it Overlaps.} Two VB states $\vert \varphi \rangle$ and $\vert \psi \rangle$ have an non vanishing overlap $\langle \varphi \vert \psi \rangle$. To compute this quantity it is convenient to consider the loop diagram obtained by superimposing both configurations (see. figure \ref{fig:overlap}). Because loops are decoupled, $\langle \varphi \vert \psi \rangle$ is the product of each loop contribution. Since there is only two ways to describe any loop with antiparallel spins, $\vert  \uparrow \downarrow \ldots \downarrow \rangle$ and $\vert  \downarrow \uparrow  \ldots \uparrow \rangle$, the overlap is  $2^{n_l}$ (up to a normalization constant) with $n_l$ the total number of loops. The normalization is fixed by  $\langle \varphi \vert \varphi \rangle = 1$ which diagram contains $N/2$ trivial loops. The
result is then $\langle \varphi \vert \psi \rangle = \epsilon_{\varphi , \psi} 2^{n_l-N/2}$ where the sign $\epsilon_{\varphi , \psi}$ is due to the relative orientations of dimers in $\vert \varphi \rangle$ and $\vert \psi \rangle$. In the case of  nearest neighbor VB on a bipartite lattice this sign can be fixed to 1 by convention, but in general this sign cannot be considered as constant. 

{\it Orthogonal states in the thermodynamic limit.} Let us consider two VB states such that the loop diagram contains at least one large loop, namely a loop involving $\alpha N^\beta$ sites with $\beta \neq 0$. The number of remaining sites is $N-\alpha N^\beta$ so the maximal total number of loops is $1+(N-\alpha N^\beta)/2$.
Hence, $\vert \langle \varphi \vert \psi \rangle \vert \leq 2^{1-(1/2)\alpha N^\beta}$ and the overlap goes to zero when $N$ goes to infinity.

Another class of orthogonal states at thermodynamic limit is formed by states whose loop diagram contains an extensive number of loops $n_l = \alpha N$ (note that $\alpha \leq 1/2$). If $\alpha < 1/2$, then $\vert \langle \varphi \vert \psi \rangle \vert \leq 2^{N(\alpha-1/2)}$ and the two states are orthogonal when $N$ goes to infinity.

\begin{figure}
\centerline{\includegraphics*[angle=0,width=\linewidth]{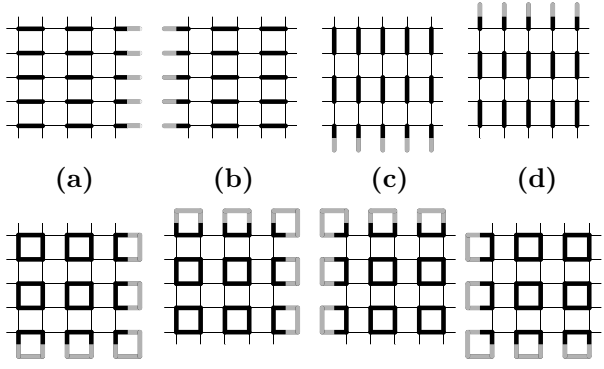}}
  \caption{\label{fig:trialstates}
Definitions of columnar (top row) and plaquette (bottom row) state.}
\end{figure}

\begin{figure}
\centerline{\includegraphics*[angle=0,width=0.8\linewidth]{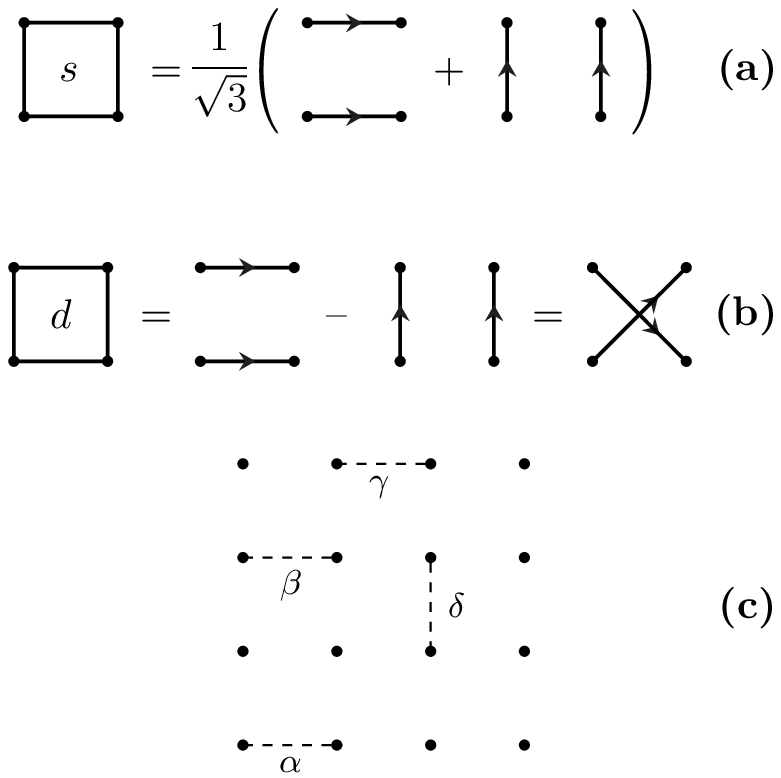}}
  \caption{\label{fig:annexe}
Definitions of {\it s-wave} plaquette {\bf (a)}, {\it d-wave} plaquette {\bf (b)} and inequivalent couples of bonds $(\alpha,\beta)$, $(\alpha,\gamma)$, $(\alpha,\delta)$ {\bf (c)}.}
\end{figure}

{\it Columnar state and plaquette states.} We define the columnar state (resp. plaquette state) as
the  equal weight linear combination of the four states (see figure \ref{fig:trialstates}) obtained by translation of
the columnar (resp. plaquette) covering of the lattice. The resulting state has a ${\bf k}=(0,0)$ momentum, thus allowing direct comparisons
with the finite size  ${\bf k}=(0,0)$ GS discussed in the article. Note that two different plaquette states can be defined on four sites~: one is symmetric
upon rotation of the plaquette (see figure \ref{fig:annexe} {\bf a}) and the second is antisymmetric (see figure \ref{fig:annexe} {\bf b}). We refer
to these states respectively as $\psi_s$ and $\psi_d$. The columnar state is denoted by $\psi_c$. 

Using the results of the previous paragraph we can show that the four components of $\psi_c$ are mutually orthogonal in the thermodynamic limit. It is easy to
check (see figure \ref{fig:trialstates}, top row) that the overlap diagram have either at least one large loop (in fact $\sim \sqrt{N}$) or an extensive number of loops. 

The very same argument can be applied for the four components of $\psi_d$ after rewriting each $d$-plaquette as crossing dimers along the diagonals (see figure \ref{fig:annexe} {\bf b}).

The case of $\psi_s$ deserve more attention. Let us define an operator $\cal U$ that change the orientation of all the dimers on half of the vertical (or horizontal) lines in an alternating pattern.
This operator is trivially self-adjoin and ${\cal U}^2 = id$, so $\cal U$ is a unitary transform and thus conserves scalar product~: $\langle \varphi \vert \psi \rangle =  \langle \varphi_{\cal U} \vert \psi_{\cal U} \rangle$ with $\vert \varphi_{\cal U}  \rangle = {\cal U} \vert \varphi \rangle$ and $\vert \psi_{\cal U}  \rangle = {\cal U} \vert \psi \rangle$. Since the action of ${\cal U}$ on a plaquette covering simply exchange  $s$-plaquettes and $d$-plaquettes (see figures \ref{fig:annexe} {\bf a} and {\bf b}), the orthogonality of the four components of $\psi_s$ is shown.

The absence of interference between components of $\psi_c, \psi_s$ or $\psi_d$  also occurs in the computation of $\langle P_\alpha \rangle$ or  $\langle P_\alpha P_\mu \rangle$ where $P_b$ denotes the operator that permutes the two sites of bond $b$ and $\mu=\beta,\gamma,\delta$ (see figure \ref{fig:annexe} {\bf b}). Indeed, the permutation of two or four sites on one component does not affect the existence of either $\sim \sqrt{N}$ large loops nor an extensive number of loops when overlapping with another component. 

{\it Dimer-dimer correlations.} The aim of this section is to compute $\langle P_\alpha P_\mu \rangle - \langle P_\alpha \rangle^2$. Note that
since $P_{i,j}= 2 {\bf S}_i .{\bf S}_j +1/2$, the correlation $\langle P_\alpha P_\mu \rangle - \langle P_\alpha \rangle^2$ is just related to the same expression with spin operators by a factor 4.

\begin{table*}[htbp]
    \centering
        \begin{tabular}{|c|c|c|c|c|c|c|c|c|c|c|c|}
        \hline \hline
    {\bf Trial state}   & \multicolumn{3}{c}{{\bf Columnar} ($\psi_c$)} & \multicolumn{4}{|c}{{\it s-wave} {\bf Plaquette} ($\psi_s$)} & \multicolumn{4}{|c|}{{\it d-wave} {\bf Plaquette} ($\psi_d$)} \\ \hline
  {\bf pairs of bonds $\rightarrow$} & $(\alpha,\beta)$ & $(\alpha,\gamma)$ & $(\alpha,\delta)$ & \multicolumn{2}{c|}{$(\alpha,\beta)$} & $(\alpha,\gamma)$ & $(\alpha,\delta)$ & \multicolumn{2}{c|}{$(\alpha,\beta)$} & $(\alpha,\gamma)$ & $(\alpha,\delta)$ \\ \hline
 {\bf (a)}        & +1         &    -1/2       & -1/2       & +1/4  &{\it +1}       & -1/4       & +1/4    & +1/4 &{\it +1}       & +1/4        & +1/4       \\
 {\bf (b)}        & +1/4       &    -1/2       & +1/4       & +1/4  &{\it +1/4}     & -1/4       & -1/4    & +1/4  &{\it +1/4}      & +1/4        & +1/4       \\
 {\bf (c)}        & +1/4       &    +1/4       & -1/2       & +1/4  &{\it +1/4}     & -1/4       & -1/4    & +1/4  &{\it +1/4}      & +1/4        & +1/4       \\
 {\bf (d)}        & +1/4       &    +1/4       & +1/4       & +1/4  &{\it +1/4}     & -1/4       & +1/4    & +1/4 &{\it +1/4}       & +1/4        & +1/4       \\ \hline
  {\bf Mean value}& {\bf 7/16} & {\bf -1/8}  & {\bf -1/8} & {\bf +1/4} & {\it +7/16} & {\bf -1/4} & {\bf 0} & {\bf +1/4} & {\it +7/16} & {\bf +1/4}  & {\bf +1/4} \\ \hline\hline
        \end{tabular}
    \caption{$\langle P_\alpha P_b \rangle - \langle P_\alpha \rangle^2$ with 
$b=\beta,\gamma,\delta$ computed in each of the four components (labelled from (a) to (d) in 
figure \protect{\ref{fig:trialstates}}) of the 3 trial states $\psi_c$, $\psi_s$ and $\psi_d$. Italic values correspond to the peculiar short range case where $(\alpha,\beta)$ share the same plaquette.}
    \label{tab:details}
\end{table*}

By a direct evaluation we derive the basic rules to compute $\langle P_\alpha \rangle$ for {\it one component} of $\psi_c, \psi_s$ or $\psi_d$~:
\begin{itemize}
\item $\langle P_\alpha \rangle_c=-1$ if $\alpha$ is occupied by a dimer, $+1/2$ otherwise,
\item $\langle P_\alpha \rangle_s=-1/2$ if $\alpha$ belongs to a plaquette, $+1/2$ otherwise,
\item $\langle P_\alpha \rangle_d=+1/2$ whatever $\alpha$ belongs or not to a plaquette.
\end{itemize}
 
Using these rules it is possible to evaluate 
$\langle P_\alpha P_\mu \rangle - \langle P_\alpha \rangle^2$  by a simple inspection of the 4 components
contributions (see figure \ref{fig:annexe}) 
of $\psi_c$, $\psi_s$ or $\psi_d$ as shown in table \ref{tab:details}. We summarize in table \ref{tab:correlationsresult}
the expected dimer-dimer correlation values in units of permutations and spin operators as well as expected
VBC and Columnar structure factors according to definition (\ref{eq:structure_factors}).  Note that
we only consider bonds $\beta$, $\gamma$ and  $\delta$ that do not share sites with $\alpha$. Also remark that for plaquette states ($\psi_s$ and $\psi_d$), very short range $(\alpha,\beta)$ correlations differ from longer range ones when $(\alpha,\beta)$ belong to the same plaquette. These short range anomalies are reported in tables \ref{tab:details} and \ref{tab:correlationsresult} in italic. 

\begin{table*}[htbp]
    \centering
        \begin{tabular}{|c|c|c|c|c|c|}
        \hline \hline
{\bf Trial state}  & {\bf  Columnar} ($\psi_c$) & \multicolumn{2}{c|}{{\it s-wave} {\bf  Plaquette} ($\psi_s$)} & \multicolumn{2}{c|}{{\it d-wave} {\bf  Plaquette} ($\psi_d$)} \\ \hline \hline
$\langle P_\alpha \rangle$ & 1/8 & \multicolumn{2}{c|}{0} & \multicolumn{2}{c|}{1/2}            \\ \hline
$\langle P_\alpha P_\beta \rangle$ & 7/16 & \multicolumn{1}{p{0.15\linewidth}|}{\centering{1/4}} & {\it 7/16} & \multicolumn{1}{p{0.15\linewidth}|}{\centering{1/4}}    & {\it 7/16}        \\
$\langle P_\alpha P_\beta \rangle - \langle P_\alpha \rangle^2$ & 27/64 & 1/4 &{\it 7/16}&  0   & {\it 3/16}        \\ 
$\langle ({\bf S}.{\bf S})_\alpha ({\bf S}.{\bf S})_\beta \rangle - \langle ({\bf S}.{\bf S})_\alpha \rangle^2$ & 27/256 & 1/16 &{\it 7/64}& 0 & {\it 3/64}\\
{\text Normalized} $(\alpha,\beta)$ & {\bf 1} & {\bf 1} &{\bf 7/4}& {\bf 0} & {\bf 3/7} \\ \hline
$\langle P_\alpha P_\gamma \rangle$ & -1/8 & \multicolumn{2}{c|}{-1/4} & \multicolumn{2}{c|}{1/4}           \\
$\langle P_\alpha P_\gamma \rangle- \langle P_\alpha \rangle^2$ & -9/64 & \multicolumn{2}{c|}{-1/4} & \multicolumn{2}{c|}{0}            \\ 
$\langle ({\bf S}.{\bf S})_\alpha ({\bf S}.{\bf S})_\beta \rangle - \langle ({\bf S}.{\bf S})_\alpha \rangle^2$ & -9/256 & \multicolumn{2}{c|}{-1/16} & \multicolumn{2}{c|}{0}\\
{\text Normalized} $(\alpha,\gamma)$& {\bf -1/3} & \multicolumn{2}{c|}{{\bf -1}} & \multicolumn{2}{c|}{{\bf 0}}             \\ \hline
$\langle P_\alpha P_\delta \rangle$ & -1/8 & \multicolumn{2}{c|}{0} & \multicolumn{2}{c|}{1/4}          \\
$\langle P_\alpha P_\delta \rangle- \langle P_\alpha \rangle^2$ & -9/64 & \multicolumn{2}{c|}{0} & \multicolumn{2}{c|}{0}           \\ 
$\langle ({\bf S}.{\bf S})_\alpha ({\bf S}.{\bf S})_\beta \rangle - \langle ({\bf S}.{\bf S})_\alpha \rangle^2$ & -9/256 & \multicolumn{2}{c|}{0} & \multicolumn{2}{c|}{0} \\
{\text Normalized} $(\alpha,\delta)$& {\bf -1/3} & \multicolumn{2}{c|}{{\bf 0}} & \multicolumn{2}{c|}{{\bf 0}}              \\ \hline \hline 
 \multicolumn{1}{|c}{\bf Correlation snapshots} & \multicolumn{1}{|m{0.2\linewidth}}{\includegraphics*[angle=0,width=\linewidth]{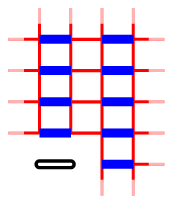}} & \multicolumn{2}{|m{0.2\linewidth}}{\includegraphics*[angle=0,width=\linewidth]{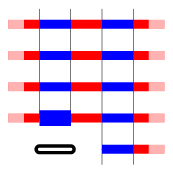}} &   \multicolumn{2}{|m{0.2\linewidth}|}{\includegraphics*[angle=0,width=\linewidth]{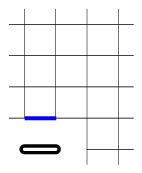}}     \\ \hline \hline 
 $C^{\infty}_{\text{VBC}}$ & 9/256 = 0.035156 & \multicolumn{2}{c|}{1/32  = 0.03125} & \multicolumn{2}{c|}{0} \\ \hline
 $C^{\infty}_{\text{col}}$ & 9/256 = 0.035156 & \multicolumn{2}{c|}{0} & \multicolumn{2}{c|}{0} \\ \hline
 $C^{\infty}_{\text{col}} / C^{\infty}_{\text{VBC}}$ & 1 & \multicolumn{2}{c|}{0} &  \multicolumn{2}{c|}{Undefined}\\ \hline\hline
        \end{tabular}
    \caption{Expectations values of correlations and structure factors for $\psi_c$, $\psi_s$ and $\psi_d$. Note that for the plaquette
    states, the correlations on the bonds next to reference one differs from the others (see italic numbers in columns 3 and 4).}
    \label{tab:correlationsresult}
\end{table*}



\begin{thebibliography}{}
\bibitem{note-ref} On a finite cluster, this degeneracy is removed and leads to quasi-degenerate
eigenstates of different quantum numbers.
\end{thebibliography}
\end{document}